\documentclass[twocolumn,showpacs,PRL,amsmath,amssymb]{revtex4}
\usepackage{graphics} 
\usepackage{graphicx} 
\usepackage{dcolumn}
\usepackage{bm}
\usepackage{epsfig}
\usepackage{epspdfconversion}
\usepackage{epstopdf}

\begin{document}

\title{Evidence for Dirac Nodes from Quantum Oscillations in SrFe$_2$As$_2$}
\author{Mike Sutherland$^{1}$, D.~J.~Hills$^{1}$,
B.~S.~Tan$^{1}$, M.~M.~Altarawneh$^{2}$, N.~Harrison$^{2}$, J.~Gillett$^{1}$, E.~C.~T.~O'Farrell$^{1}$, T.~M.~ Benseman$^{1}$, I.~Kokanovic$^{3}$, P.~Syers$^{1}$, J.~R.~Cooper$^{1}$ and Suchitra~E.~Sebastian$^{1}$}

\affiliation{$^1$Cavendish Laboratory, University of Cambridge, J.J. Thomson Ave, Cambridge, CB3 0HE UK}
\affiliation{$^2$Los Alamos National Laboratory, Los Alamos, New Mexico, 87545 USA}
\affiliation{$^3$University of Zagreb, Faculty of Sciences, Department of Physics, Zagreb, Croatia}


\begin{abstract}
 
We present a detailed study of quantum oscillations in the antiferromagnetically ordered pnictide compound SrFe$_2$As$_2$ as the angle between the applied magnetic field and crystalline axes is varied. Our measurements were performed on high quality single crystals in a superconducting magnet, and in pulsed magnetic fields up to 60~T, allowing us to observe orbits from several small Fermi surface pockets. We extract the cyclotron effective mass $m^{\star}$ and frequency $F$ for these orbits and track their values as the field is rotated away from the $c$-axis. While a constant ratio of $m^{\star}/F$ is observed for one orbit as expected for a parabolic band, a clear deviation is observed for another. We conclude that this deviation points to an orbit derived from a band with Dirac dispersion near the Fermi level.

 \end{abstract}


\maketitle

Electrons with Dirac-like dispersion have been invoked to explain many fascinating phenomena in materials. Arising in systems where the low-energy states are governed by the relativistic Dirac equation, they are characterized by an energy which depends linearly on wavevector, and possess high velocities and a zero effective mass. Dirac electrons determine the remarkable electronic properties of graphene \cite{Novoselov05}, and in topological insulators they may give rise to novel spin textured states through strong spin-orbit coupling \cite{Xu11}. Recent experimental work has established a compelling body of evidence for the existence of such electrons - in graphene \cite{Neto09,Gruneis09} and graphite \cite{Zhou06} for instance, and in topological insulators such as Bi$_2$Se$_3$ \cite{Zhang09}.   

In this letter we report evidence from quantum oscillation experiments for the existence of Dirac fermions arising in SrFe$_2$As$_2$, an antiferromagnetically ordered metal which becomes superconducting by the application of hydrostatic pressure \cite{Alireza09,Kotegawa09} or by chemical substitution \cite{Leithe-Jasper08}.  SrFe$_2$As$_2$ possesses a Fermi surface of several small pockets, formed due to the reconstruction of a larger paramagnetic Fermi surface by magnetic ordering in the FeAs planes along the \textbf{Q}=(\textit{$\pi,\pi,0$}) direction \cite{Analytis09}. Our study tracks the effective mass m$^\star$ of cyclotron orbits  \cite{Shoenberg84} with frequency $F$ that arise from these pockets, as the angle between the applied magnetic field and crystalline axes is varied. When the orbits enter the polar regions of one Fermi surface, we observe a decrease in the ratio $m^{\star}/F$ from the value expected for a parabolic band.

We successfully model this decrease in terms of a Dirac dispersion at the Fermi level, arising in an effective two-band model parameterized by only the interlayer hopping $t$ and chemical potential $\mu$. This approach was suggested previously \cite{Harrison09b}, but until now complete angular data was not available to confirm it. The broad agreement between our measurements and this simple model establishes the existence of massless excitations and constrains their properties.

Quantum oscillatory studies offer an extremely sensitive bulk probe of electronic structure. Unlike transport measurements, which take into account a weighted average of contributions from various bands, quantum oscillation studies are able to selectively access information about each band individually. The presence of such particles in our measurements strongly suggests a topologically protected nodal spin density wave state \cite{Ran09}, which occurs due to a crossing of hole and electron bands near the Fermi level resulting from the magnetic reconstruction of the Fermi surface. Intriguingly, Dirac fermions arising in this manner are proposed to be fundamentally different than those in graphene. In the pnictide case, the presence of non-Dirac electrons leads to two Dirac cones with the same chirality, with distinct implications for transport properties \cite{Morinari10}. 


Our samples were grown using a self-flux technique \cite{Sebastian08},
yielding high quality platelet-like single crystals with dimensions of
approximately 1.5~$\times$ 1.5~$\times$ $0.05$~mm$^3$ that were free of
flux inclusions. To improve sample quality, single crystals were
annealed at 800$^\circ$C for 24 hours under argon gas, with subsequent
annealing for 2 hours at 300$^\circ$C in some instances to remove a
superconducting signal near 20~K \cite{Saha09} without otherwise
altering the  $\rho_{T}$ curve. Annealed single crystals were of high
quality, with $\rho_{300K}/\rho_{4K} $ values of up to 17 being reached,
Over 20 annealed samples were then screened for best quantum
oscillations. Quantum oscillations at low fields down to 12T (Fig. 1a)
were measured for a crystal with $\rho_{300K}/\rho_{4K}  \approx$ 9.


Oscillations were studied using two techniques, via resistivity measurements conducted on a superconducting cryomagnetic system in Cambridge, with fields as high as 16~T, and with a contactless tunnel diode oscillator (TDO) technique \cite{Sebastian08} using the 60~T short pulse magnet at Los Alamos National Laboratory at temperatures from 0.5 K to 5 K. In both systems we are able to vary the orientation of the crystal by use of a mechanical rotator.

       \begin{figure} \centering
              \resizebox{8.5cm}{!}{
              \includegraphics[angle=0]{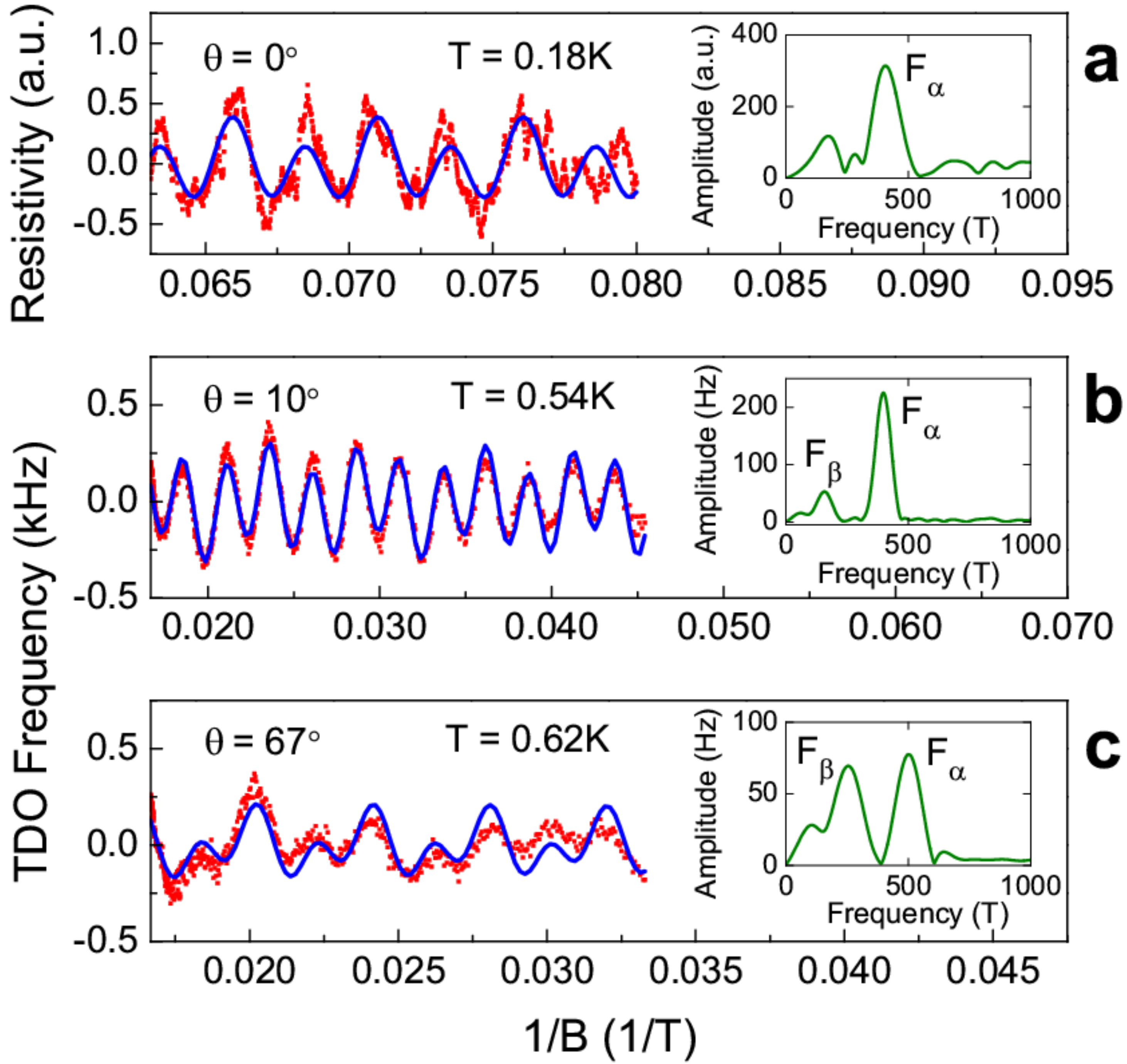}}
              \caption{\label{fig:1}(color online) Quantum oscillations in SrFe$_2$As$_2$, plotted against inverse field. Panel (a) shows resistivity data taken in a superconducting magnet, while panels b) and c) show the oscillatory component of the resonance frequency of a TDO circuit in a pulsed magnetic field. Data is shown for different orientations of the field to the crystalline $c$ axis, defined by $\theta$. The solid lines are a fit to a two-frequency model, described in the text. The insets show the Fourier spectra, dominated by two peaks which we identify as $F_{\alpha}$ and $F_{\beta}$.}
              
      \end{figure}
    
Figure 1 shows examples of our data, with $\theta$ defined as the angle between the applied magnetic field and the crystalline $c$-axis. The top panel shows results of a four-wire resistivity measurement in a 16~T superconducting magnet, while the bottom panels show the resonance frequency of a TDO circuit which was coupled inductively to a sample subjected to a pulsed magnetic field. In both cases oscillations arise due to changes in the magnetoresistance of the crystal, described by the Shubnikov-de Haas (SdH) effect.  A fourth order polynomial has been used to subtract the field-dependent background, which arises due to the magnetoresistance of the sample. Our data is plotted in units of $1/B$, indicating an oscillatory component of the signal that is periodic in inverse field. 

At the lowest angles, we observe oscillations down to B $\sim$ 12 Tesla, whose amplitude is exponentially dependent on field. The data is well described by a simple fit to a model with two frequencies, $F_\alpha$ and $F_\beta$, with amplitudes $A$ and phases $\delta$. The total oscillation amplitude is then $\tilde{A} = A_\alpha sin(2\pi F_\alpha/B+\delta_\alpha)+A_\beta sin(2\pi F_\beta/B+\delta_\beta)$, and the results of this fit are shown by the solid lines through the data.  A Fourier transform of each data set is shown in the insets of Fig. 1, showing a spectrum dominated by two frequencies which are evident at all angles. 

We identify these frequencies as corresponding to orbits about small pockets of the reconstructed Fermi surface. In keeping with convention for the pnictide `122' compounds \cite{Sebastian08,Analytis09,Harrison09} we label these as $\beta$ and $\alpha$ for the lower and higher frequencies respectively,  with $F_\beta$ = 160 T and $F_\alpha$ = 400 T for $\theta$=10$^{\circ}$. At higher angles the overall amplitude of the oscillations is reduced, however we are still able to resolve the $\alpha$ and $\beta$ frequencies. For the highest angles, a third frequency, $F$=200 T at $\theta$ = 74$^{\circ}$ becomes resolvable, which we ascribe to the $\gamma$ orbit.

The magnetic field dependence of the amplitude for the $\theta$=10$^{\circ}$ data yields Dingle temperatures of $T_{D\alpha}$ = 0.7 K and $T_{D\beta}$ = 2.7 K. From these we deduce mean free paths of 2700 \AA \ and 800 \AA \ respectively, using Fermi surface averaged velocities of $v_{\alpha} = \sqrt{2\hbar e F_{\alpha}}/m^\star$ = 5 $\times$ 10$^4$ ms$^{-1}$ and $v_{\beta}$ = 6 $\times$ 10$^4$ ms$^{-1}$. These long mean free paths are comparable to those reported in BaFe$_2$As$_2$ \cite{Analytis09}.


       \begin{figure} \centering
              \resizebox{8.5cm}{!}{
              \includegraphics[angle=0]{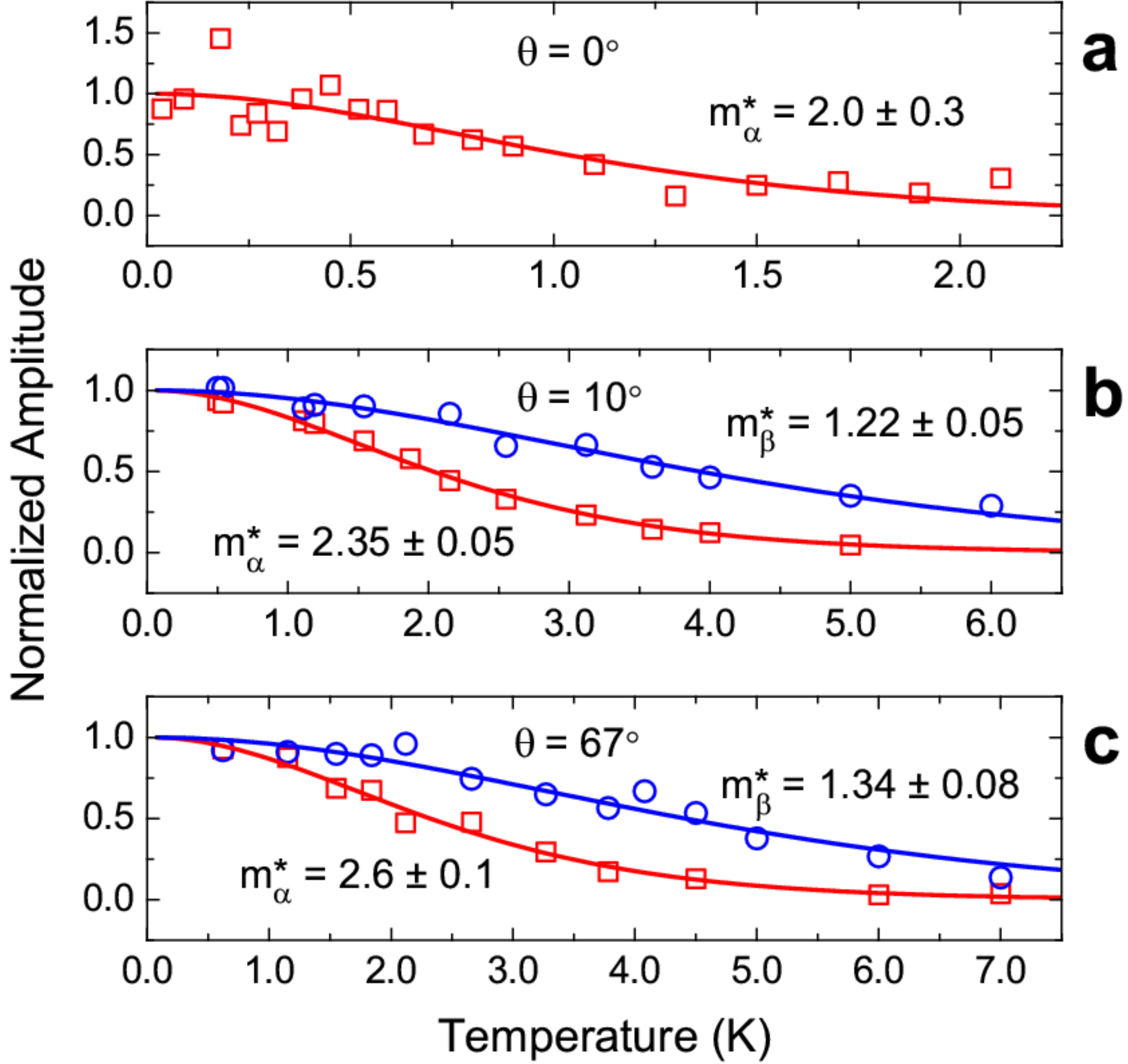}} 
              \caption{\label{fig:2}(color online) Amplitude of oscillations as a function of temperature, defined by the height of the peaks in the Fourier spectrum. The amplitudes are normalized to the low temperature values. The lines through the data represent fits to the Lifshitz-Kosevich damping factor used to extract m$^{\star}$.}
      \end{figure}
      
In Fig. 2 we show how the amplitude of the oscillations depends on temperature. We plot the height of the peak in the Fourier spectrum as a function of temperature, normalized to the low temperature value. This data is then fit with the Lifshitz-Kosevich (LK) thermal damping factor, $R_T$ = $X/sinh(X)$ where $X$=14.69 $m^{\star}T/B$ \cite{Shoenberg84}. For the high field data set, we worked with Fourier windows between 22 and 60~T, and define $B^{-1}$ as the average inverse field taken over the width of a window. For the low field data set the Fourier window was 12.5-16~T. 

For both the $\alpha$ and $\beta$ frequencies good quality fits are obtained for high and low angles, with effective masses m$^{\star}_{\alpha}$ = 2.35 $\pm$ 0.05 and m$^{\star}_{\beta}$ = 1.22 $\pm$ 0.05 at $\theta$ = 10$^\circ$, expressed in units of $m_e$ and with error bars set by uncertainties in the least squares fit. The top panel shows data taken in the superconducting magnet system, giving a value for $m^{\star}_\alpha$ of 2.0 $\pm$ 0.3 at $\theta$ = 0$^\circ$, which is in agreement with the high field data, serving as a useful check of consistency between measurement techniques and systems. In panel (c) data at higher angles is shown, demonstrating an increase in $m^{\star}$ for both orbits \cite{Comment1}. 

       \begin{figure} \centering
              \resizebox{8.5cm}{!}{
              \includegraphics[angle=0]{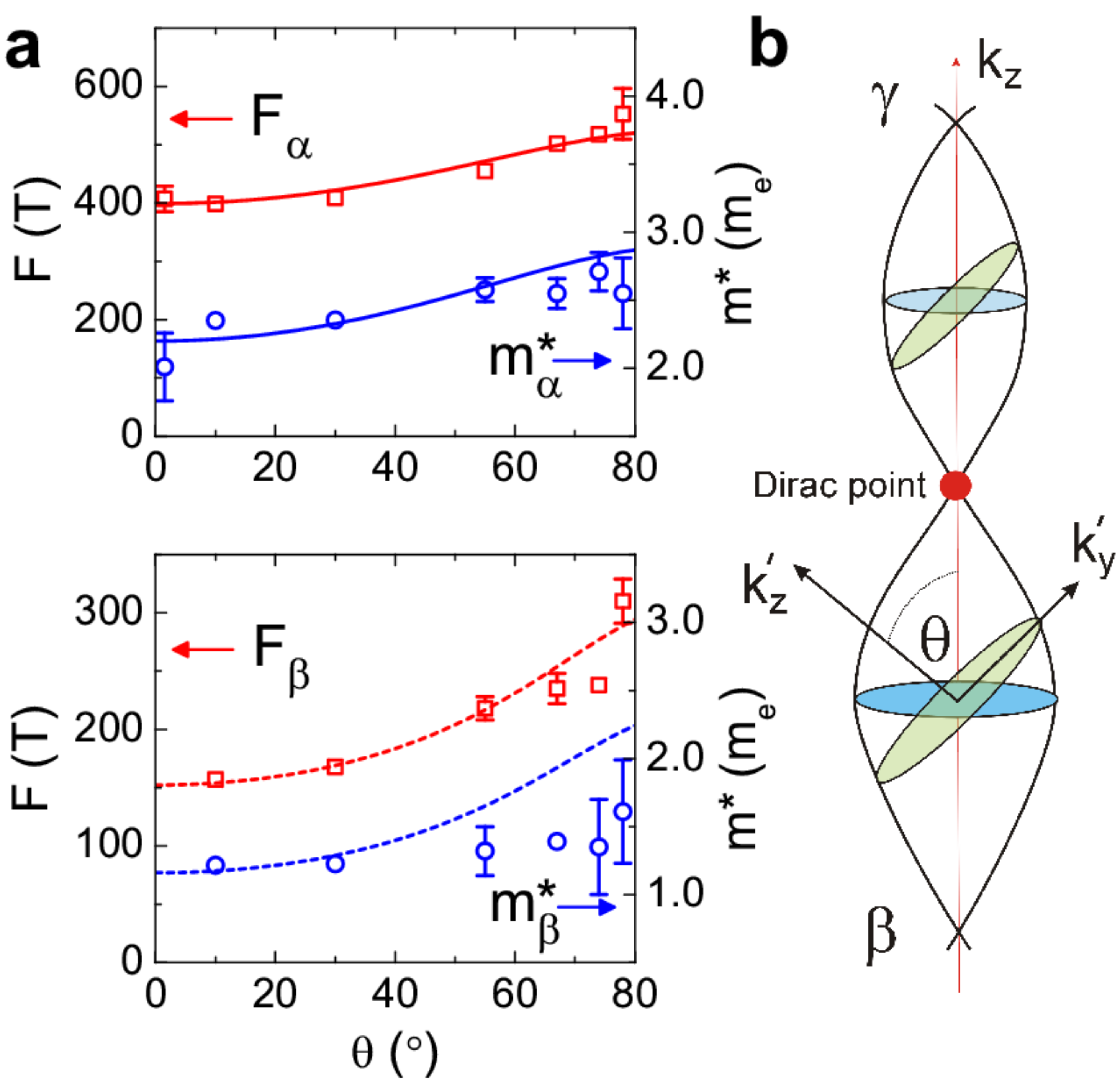}}
              \caption{\label{fig:3}(color online) (a) The upper panel shows frequency ($F$) and effective mass ($m^{\star}$) as a function of angle for the $\alpha$ electron pocket. The lines are fits assuming conventional orbits about an ellipsoid formed from a parabolic band. The lower panel shows the same data for the $\beta$ orbit. (b) Cartoon showing extremal orbits of the Fermi surface near the Dirac point arising from Eq. \ref{eq:k}. $k_y^{'}$, and $k_{x}$ lie in the plane of the orbit, while $k_z$ lies along the $c^{'}$ axis of the reciprocal lattice. The $\beta$ ($\gamma$) orbit originates from the larger (smaller) pocket, while the $\alpha$ orbit encompasses a separate Fermi pocket derived from parabolic bands (not shown).}
              
      \end{figure}
     
      
We collect $m^\star$ and $F$ for all angles together in Fig. 3, with fitting errors reflecting the signal to noise, which is limited at higher angles by the shorter field range of the oscillations.  The top panel shows data for the $\alpha$ orbit, which shows that the increase in $F_\alpha$ is closely mirrored by a concomitant increase in $m^\star_\alpha$. Intriguingly, the same cannot be said for the $\beta$ orbit in the bottom panel, where $F_\beta$ rises faster with angle than  $m^\star_\beta$. It is from these observations that the presence of Dirac fermions in SrFe$_2$As$_2$ can be inferred.

To interpret our data we now consider two different scenarios. First, that the small Fermi surface pockets are ellipsoids derived from conventional parabolic bands, and second, that the pockets originate from bands with a linear dispersion near $E_F$. In the first case, we would expect the frequency as a function of angle to be given by \cite{Shoenberg84}:

 \begin{equation}
 F(\theta) = \frac{m^\star_0 \mu}{e\hbar \sqrt{cos^2\theta+\xi^2sin^2\theta}}
 \label{eq:dispersion2}
 \end{equation}
 
 
 Where $\mu$ is the chemical potential, $\xi$ is the ratio of major to minor axes of the extremal elliptical cross section, and $m^\star_0$ is the effective mass measured at $\theta = 0$. $m^{\star}(\theta)$ has the same $\theta$ dependence as $F(\theta)$, with $m^{\star}(\theta)$ = ($e\hbar/\mu)F(\theta)$. Fig. 3a shows the results of a best fit to this model. While a fit with parameters with $\mu$ = 21 meV, $m^\star_0$=2.2 and $\xi$=0.58 captures the angular dependence of the frequency and mass for the $\alpha$ orbit, it fails to explain the experimental data for the $\beta$ orbit. The best fit to $F_\beta(\theta)$ yields $\mu$ = 15 meV, $m^\star_0$=1.2 and $\xi$=0.25, which rises too fast with angle to account for the angular dependence of $m^\star_\beta$. We are thus forced to look beyond a conventional model to describe this pocket.
 
In the 122-pnictide materials, Dirac points are proposed to arise due to degenerate electron and hole bands that meet at $E_F$. The existence of a small amount of interlayer hopping provides access to these points, by creating a series of stacked pockets along the $k_Z$-axis that touch at their extrema. The influence of massless charge carriers originating from these regions of the Fermi surface on quantum oscillatory phenomena has been captured in a minimal model \cite{Harrison09b}, which predicts a dip in the ratio $m^\star(\theta)/F(\theta)$ at higher angles consistent with our observation for the $\beta$ orbit. The starting point is a dispersion relation of the form:

 \begin{equation}
 \epsilon = \pm \hbar v^\star |k| + 2 t cos\left(ck_z/2\right)+\mu
 \label{eq:dispersion}
 \end{equation}
  
\noindent where $c$ is the bilayer spacing for the body centered tetragonal crystal structure (12.3 \AA), $v^\star$ is the characteristic Dirac velocity, $t$ is the interlayer hopping parameter, $|k|=\sqrt{k_x^2+k_y^2}$, and $\mu$ is the chemical potential. For simplicity, the characteristic velocity $v^\star$ is taken to be isotropic, although there is evidence for a small anisotropy from angle-resolved photoemission (ARPES) measurements \cite{Richard10}. As the angle $\theta$ between the $c$-axis and the applied magnetic field is varied, the extremal orbits take approximately elliptical forms. Defining $k_x$ and $k_y'$ as vectors lying in the planes of these orbits, as illustrated in Fig. 3b, gives

 \begin{equation}
 k_x^{\mp} = \sqrt{\left(\frac{2t}{\hbar v^\star}cos\left[\frac{ck_y^{'}sin \theta}{2}\right]-\frac{\epsilon \mp \mu}{\hbar v^\star}\right)^2 -k_y^{'2}cos^2\theta} 
  \label{eq:k}
 \end{equation}
 
 This allows calculation of the expected angular variation of the cyclotron effective mass and extremal cross sectional areas $A_{k, \theta}$, which are directly related to the quantum oscillation frequencies via the Onsager relation, $F_{\theta}=\left(\hbar/2\pi e\right)A_{k, \theta}$. 
  
 
       \begin{figure} \centering
              \resizebox{8.5cm}{!}{
              \includegraphics[angle=0]{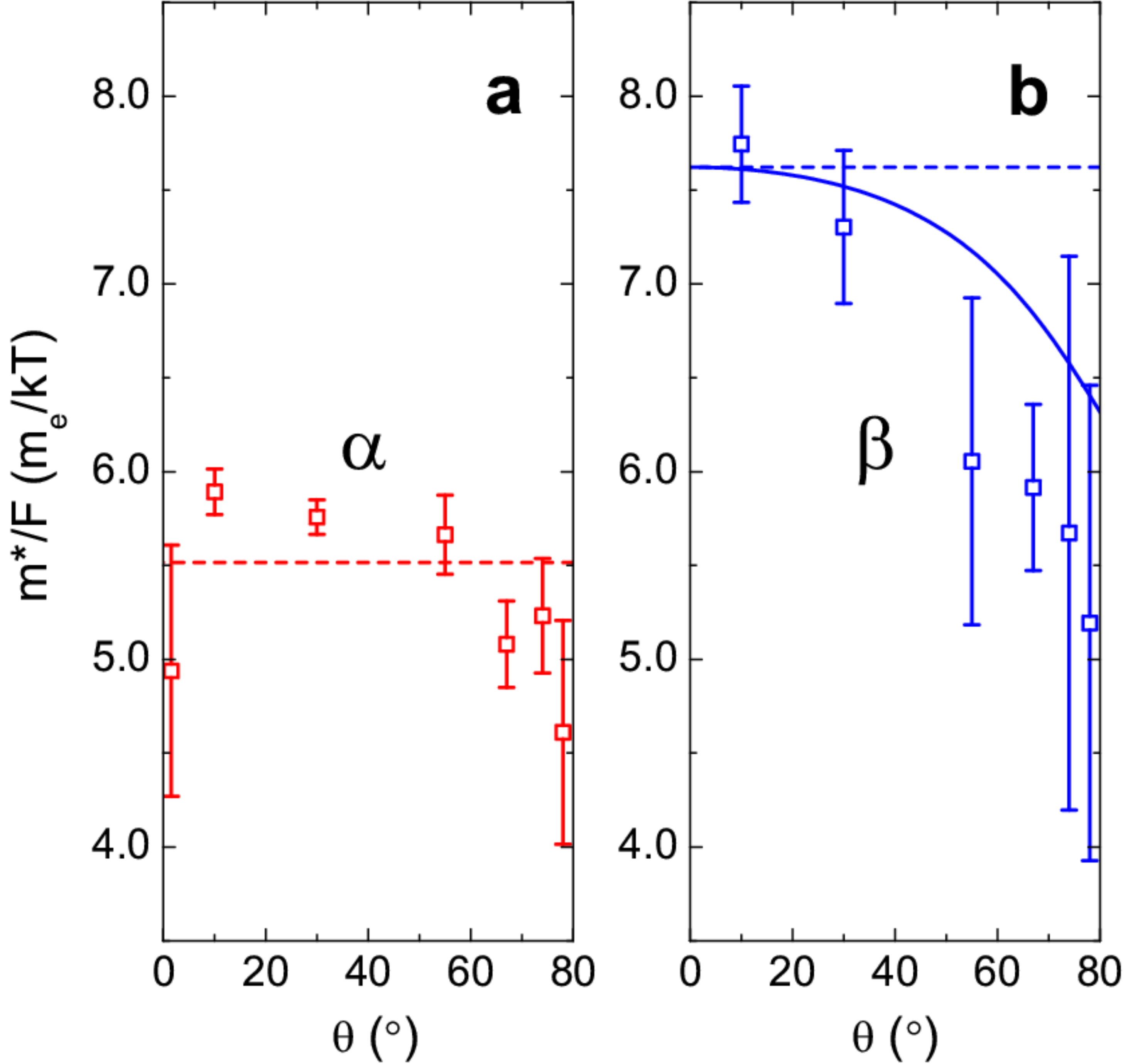}}
              \caption{\label{fig:4}(color online) Ratio of the effective mass to cyclotron frequency as a function of angle for the $\alpha$ and $\beta$ orbits. The solid line shows the fit to a model arising from the Dirac-like dispersion relation in Eq. \ref{eq:dispersion}, while the dotted line shows the expected behavior for conventional ellipsoids.}
              
      \end{figure}

Taking the ratio $m^\star(\theta)/F(\theta)$ provides a straightforward test of the predictions of this model. For Fermi surface pockets originating from bands with parabolic dispersion, we would expect a ratio which does \textit{not} vary with angle, as has for example been shown in 2D organic metals \cite{Wosnitza91}. For bands with the dispersion parameterized by Eq. \ref{eq:dispersion} a striking effect occurs - as $\theta$ is increased the orbits begin to encompass regions of the Fermi surface close to the Dirac point, where velocities are high and masses very light \cite{Harrison09}. Thus, the orbitally averaged cyclotron masses are \textit{smaller} than those expected from parabolic bands, and the ratio $m^{\star}/F$ is correspondingly reduced, providing an explanation for observation of the same. 
   
Figure 4 confirms this effect for the $\beta$ orbit. In panel (a) $m^{\star}_{\alpha}/F_{\alpha}$ is shown to be nearly constant over a wide range of angles, within the errors of the experiment, as expected for a Fermi surface with conventional parabolic bands.  Panel (b) shows the same ratio for the $\beta$ orbit, where a dip in $m^{\star}_{\beta}/F_{\beta}$ is clearly observed, with the solid line representing a fit to a Fermi surface parameterized by Eq.\ref{eq:k} with $v^\star_\beta=6.75 \times 10^4$ ms$^{-1}, t$ = 15.2 meV and $\mu \approx 0$ meV. We conclude that this is direct proof of the presence of Dirac nodes sampled by the $\beta$ orbit \cite{Comment2}.


Despite the relative simplicity of the Fermi surface model derived from Eq.\ref{eq:k}, the extracted $v^\star_\beta$ agrees rather well with that measured with ARPES \cite{Richard10,Kim11}, where a linear dispersion of bands in the vicinity of the Brillouin zone M point is observed. Through our observation of the angular dependence of quantum oscillations, we can unambiguously confirm the observation of linearly dispersing bands that cross the Fermi energy near a Dirac node. Recent evidence from magnetotransport measurements on the same material interpret a linear magnetoresistance at intermediate fields as evidence of Dirac excitations \cite{Huynh11}, which is also consistent with our findings.

Having established the presence of Dirac nodes in the parent compounds of pnictide high T$_c$ superconductors, an interesting question remains as to the role they play in superconductivity in this system, and their influence across the doping or pressure phase diagram. Recent theoretical work has suggested that the physical symmetry of the 122-pnictides and the topology of their band structure lead to an unusual robustness of the nodal spin density wave state \cite{Ran09}, with important consequences for the nature of the unconventional superconductivity induced by pressure or doping. 

We thank S. Goh, G. Lonzarich, A. Carrington and A. Coldea for discussions. M.S. and S.E.S acknowledge the Royal Society. This research was funded by the EPSRC, I2CAM and the DOE BES project ``Science at 100~T"


\begin{thebibliography}{23}
\bibitem[{Com({\natexlab{a}})}]{Comment1}
\bibinfo{note}{The low frequencies necessitate a large Fourier transform window
  in order to properly resolve the spectrum, which may lead to an underestimate
  of m$^\star$ through the field dependence of the Dingle factor. While we do
  not account for this explicitly in our analysis, simulations of our data
  suggest the masses differ by no more than 5-10 \% from their actual values.}
\bibitem[{Com({\natexlab{b}})}]{Comment2}
\bibinfo{note}{The model parameterized by Eq. 3 is the minimal one necessary to
  capture the effects of stacked Dirac nodes predicted by bandstructure. We
  expect better fits to $F_{\beta}(\theta)$ and $m^{\star}_\beta(\theta)$ would
  be obtained by a more sophisticated model including anisotropy and higher
  harmonic terms.}
\end{thebibliography}


\end{document}